\documentclass[aps,showpacs,twocolumn,superscriptaddress,prb,floatfix]{revtex4-1}
\usepackage{graphicx,color,epsfig,rotate}
\usepackage{amssymb,amsmath,bm}

\usepackage{dcolumn}  % align table columns on decimal point
\usepackage{hyperref}% add hypertext capabilities

\newcommand{\ka}{_{\bf k}}
\newcommand{\sumk}{\sum_{{\bf k}}}
\newcommand{\g} \textbf{}
\newcommand{\da}{^\dagger}

\begin{document}
\title{Dynamical structure factor of quasi-2D antiferromagnet in high fields}

%--------------------------------------------------------------------
\author{W. T. Fuhrman}
\affiliation{Department of Physics, University of California, Irvine,
California 92697, USA}
\author{M. Mourigal}
\affiliation{Institute for Quantum Matter and Department of Physics and Astronomy, 
	Johns Hopkins University, Baltimore, MD 21218, USA}
\author{M. E. Zhitomirsky}
\affiliation{Service de Physique Statistique, Magn\'etisme et
Supraconductivit\'e, UMR-E9001 CEA-INAC/UJF, 17 rue des Martyrs,
38054 Grenoble Cedex 9, France}
\author{A. L. Chernyshev}
\affiliation{Department of Physics, University of California, Irvine,
California 92697, USA}
%--------------------------------------------------------------------

\date{\today}

\begin{abstract}
We study high-field magnon dynamics and examine the dynamical structure factor in the 
quasi-2D tetragonal Heisenberg antiferromagnet with interlayer coupling corresponding 
to realistic materials. Within spin-wave theory, we show that a non-zero interlayer coupling 
mitigates singular corrections to the excitation spectrum occurring in the high-field regime 
that would otherwise require a self-consistent approach beyond the $1/S$ approximation.  
For the fields between the threshold for decays and saturation field
we observe widening of the two-magnon sidebands with significant shifting of the 
spectral weight away from the quasiparticle peak. We find spectrum broadening 
throughout large regions of the Brillouin zone, dramatic redistributions of spectral weight to 
the two-magnon continuum, two-peak structures and other features clearly 
unlike conventional single-particle peaks.
\end{abstract}
%--------------------------------------------------------------------
%--------------------------------------------------------------------
\maketitle

\section{Introduction}

Studies of quantum antiferromagnets are prominent in magnetism, elucidating
the role of symmetry, fluctuations and dimensionality in the ground state of 
many-body systems, revealing complex dynamical properties of spin excitations
and allowing detailed comparison between theory and
experiments.\cite{Manousakis,Mattis,Diep,Johnston,Balents,Giamarchi08}
A better understanding of many phenomena, from superfluidity to complex quantum phases
may also be achieved through comparative investigations of systems of strongly interacting bosons and 
frustrated antiferromagnets. \cite{,Balents,Giamarchi08,Stone06} 
 Recent developments in the synthesis of molecular based 
 antiferromagnets with moderate exchange
 constants~\cite{Woodward02,Lancaster07,Coomer07,Xiao09} 
 has opened previously unreachable high magnetic 
 field regime to experimental investigations.\cite{Tsyrulin09,Tsyrulin10}
 Furthermore,  the recent completion of neutron scattering instruments with 
 enhanced resolution~\cite{Ollivier2010,Bewley2011,Ehlers2011}
provides an opportunity to study the dynamics of quantum antiferromagnets in a much
wider momentum-energy space and under the influence of  applied magnetic field.

In a collinear antiferromagnet, e.g., a square-lattice Heisenberg antiferromagnet, magnetic field induces 
a non-collinearity of the magnetic order, similar to the effect of geometric 
frustration due to competing interactions, present in some other spin systems. \cite{J1J2J3,Henley}
Generally, non-collinearity leads to an enhanced interaction among spin 
excitations\cite{Chernyshev06,triangle} and  results in stark differences from conventional theory 
for magnons in the high-field regime.\cite{field,Mourigal2010,Kreisel08,Syromyat09}
In increasing field, spins gradually cant toward the field direction until they reach the saturation 
field, $H_s$, where quantum fluctuations are fully suppressed and ferromagnetic alignment achieved.  
For fields above a certain threshold value, $H^*$, but below saturation, coupling of the transverse and 
longitudinal spin fluctuations provides a channel for decays, through cubic terms in the 
spin-wave expansion.  In this regime, magnons are  predicted to be strongly damped,
% and non-trivial features to arise
 resulting in the loss of well-defined quasiparticle peaks.\cite{field}

Magnon decays and spectrum broadening above the threshold field 
in the purely two-dimensional square-lattice antiferromagnet have been confirmed 
via quantum Monte Carlo\cite{Olav08} (QMC) and exact diagonalization\cite{Lauchli09} 
numerical studies. In addition, QMC has revealed  a non-trivial redistribution of spectral weight 
resulting in non-Lorentzian ``double peak'' features in the dynamical structure factor, also 
previously observed in the original analytical study.\cite{field}
The inelastic neutron-scattering experiment on the spin-$5/2$ material
 $\rm{Ba_2MnGe_2O_7}$ are also indicative of the field-induced magnon decays,\cite{Masuda10}
 although not fully conclusive.\cite{Mourigal2010} 
 It can  be argued that recent thermal conductivity experiments in the so-called Bose-Einstein condensed
magnets that observed suppression of heat current in the vicinity of critical fields\cite{Kohama2011} 
are related to magnon decay dynamics.\cite{sasha_unpub} 
Current advances in neutron scattering instrumentation\cite{Ollivier2010,Bewley2011,Ehlers2011} 
open avenues to search for decays in large portions of the (${\bf k},\omega$) space. 
This is a particularly important issue in the case of spin-$1/2$ systems 
for which a comparison of experimental findings with existing theories is still missing.

In this work, we extend the previous work of three of us and
provide a theoretical investigation of high-field dynamics in the 
quasi-2D tetragonal $S\!=\!1/2$ Heisenberg antiferromagnet with interlayer coupling corresponding 
to realistic materials. In particular, within spin-wave theory in the $1/S$ approximation,
we obtain quantitative predictions for the dynamical structure factor, $S({\bf k},\omega)$, 
the quantity measured in inelastic neutron scattering experiments, for a representative
interlayer coupling ratio $J'/J\!=\!0.2$, relevant, e.g., for $\rm{(5CAP)_2 CuCl_4}$,\cite{Coomer07}
along several representative paths in the Brillouin zone and for magnetic fields of
 $H\!=\!0.90H_s$ and $H\!=\!0.95H_s$.

Within the framework of spin-wave theory,  3D coupling also  
helps with the following technical issues.
First, in the case of the 2D, $S\!=\!1/2$ square-lattice antiferromagnet in high fields,  
calculations without the self-consistent treatment of cubic magnon interactions lead to
renormalized quasiparticle peaks that unrealistically escape the (unrenormalized) 
two-magnon continuum, precluding an accurate determination of the dynamical 
structure factor.\cite{field} Second, for the same 2D case, 
the standard $1/S$ expansion for the magnon spectrum breaks down in the high-field regime, 
owing to the transfer of the van Hove singularities from the two-magnon continuum 
to the one-magnon mode due to a coupling between the two.\cite{field,Mourigal2010}
To regularize such unphysical singularities, two self-consistent schemes were developed 
 in Refs.~\onlinecite{field} and \onlinecite{Mourigal2010}.
While the former approach\cite{field} did take into account the renormalization 
of the spectrum and, as a consequence, the spectral weight redistribution, it was only partially self-consistent.
The latter, on the other hand, ignored the real part of the spectrum renormalization,\cite{Mourigal2010} 
essentially enforcing Lorentzian shapes of the quasiparticle peaks and excluding more complex 
profiles such as ``double-peak'' features.\cite{field,Olav08}  
While this latter approach is suitable for spins $S\!\geq\!1$,\cite{Mourigal2010}  
it cannot be deemed satisfactory in the case of $S\!=\!1/2$.

The aim of this work is to demonstrate that a non-zero 3D interlayer coupling largely mitigates
 the singularities of high-field corrections and thus provides a physical representation 
 of the excitation spectrum {\it without} the use of self-consistency. 
While the quasiparticle energy shift and corresponding 
escape from the ``bare'' continuum are still present to some degree within this approach, they are 
considerably weaker than in the purely 2D case, allowing for a detailed analysis of the effects of broadening
and weight redistribution in the spectrum.
Altogether, the relatively simple $1/S$ approximation requires considerably less computation than 
self-consistent methods, follows the spirit of the regular spin-wave expansion 
and allows quantitative examination of the dynamical structure factor.
This approach is applicable to all spin values including the spin-$1/2$ case, 
which is the focus of our study.  Since all real antiferromagnets are at best quasi-two-dimensional, 
our analysis is relevant to most realistic materials, even those with weak interlayer coupling. 
Results of the present work therefore consist in detailed and clear predictions for the dynamical structure 
 factor to guide experiments and a comparison with self-consistent techniques for the purely 2D case.  

The paper is organized as follows: Section II contains an overview of the spin-wave 
formalism, with full details in Appendix A.  Following in Section III is a brief discussion of decay
conditions and singularities.  Section IV contains the dynamical structure factor calculated in the 
$1/S$ approximation for non-zero interlayer coupling.  Section V contains our conclusions.  
Appendix A contains details of the $1/S$ spin-wave theory for the quasi-2D tetragonal Heisenberg
antiferromagnet in external field.  

%Appendix B shows on-shell spectrum and damping.}

% --------------------------------------------------------------------
% --------------------------------------------------------------------
\section{Interacting Spin Waves in Field}
% --------------------------------------------------------------------
% --------------------------------------------------------------------

We begin with the Heisenberg Hamiltonian of nearest-neighbor 
interacting spins on a tetragonal lattice in the presence of an applied magnetic field directed
along $z_0$ axis in the laboratory reference frame,
\begin{eqnarray}
\hat{\cal H} = \sum_{\langle ij \rangle} J_{ij}\ \mathbf{S}_i\cdot\mathbf{S}_j
		   - H \sum_{i}  S_i^{z_0} \ ,
\label{eq:haf}
\end{eqnarray}
where in-plane coupling is $J_{ij}\!=\!J$ and the interplane one is  $J_{ij}\!=\!J'\!=\!\alpha J$ and 
we assume $0\!\leq\!\alpha\!\leq\! 1$.
With the details of the technical approach explicated in Appendix~\ref{ap1}  and 
Refs.~\onlinecite{Nikuni98,field,Mourigal2010}, we summarize here the key 
steps of the spin-wave theory approach to this problem.  
% --------------------------------------------------------------------
\begin{figure}[t]
\includegraphics[width=0.90\columnwidth]{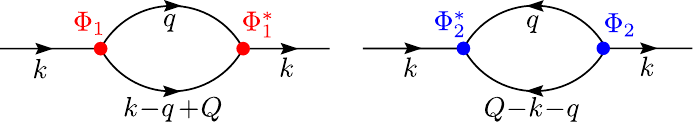}
\caption{(Color online) Self-energies in the lowest $1/S$ order obtained from decay (left) and source (right) interactions.}
\label{fig1}
\end{figure}
% --------------------------------------------------------------------

First, we identify the canted spin configuration in the equivalent {\it classical} spin model. 
Then we quantize the spin components in the rotating frame that aligns the local spin 
quantization axis on each site in the direction given by such a classical configuration. 
Application of the standard Holstein-Primakoff transformation bosonizes spin operators.  
After subsequent Fourier transformation, the Hamiltonian is diagonalized via the 
Bogoliubov transformation, yielding the  Hamiltonian that can be written as\cite{triangle}
\begin{eqnarray}
\hat{\cal H} = \sum_{\bf k} \tilde{\varepsilon}_{\bf k} 
                     b_{\bf k }^{\dagger} b_{\bf k}^{\phantom{\dagger}}
  &+& \frac{1}{2!} \sum_{\bf k, q}  \Phi_{1}({\bf k, q}) 
                     \bigl( b_{\bf k-q+Q}^{\dagger} b_{\bf q }^{\dagger} 
                            b_{\bf k }^{\phantom{\dagger}} + \rm{h.c.}         
                     \bigr) \label{eq:cubic} \\ 
&+&  \frac{1}{3!} \sum_{\bf k, q}  \Phi_{2}({\bf k, q}) 
                     \bigl( b_{\bf Q-k-q}^{\dagger} b_{\bf q }^{\dagger} 
                            b_{\bf k }^{\dagger} + \rm{h.c.}
                     \bigr) + ...
\nonumber
\end{eqnarray}
where the ordering vector ${\bf Q}=(\pi,\pi,\pi)$ enters the momentum conservation condition 
because of the staggered canting of spins.\cite{Mourigal2010} 
In this expression, $\tilde{\varepsilon}_{\bf k} = \varepsilon_{\bf k} + \delta\varepsilon_{\bf k}$, where
 $\varepsilon_{\bf k}$ is the ``bare'' magnon dispersion given by linear spin-wave theory and
  $\delta\varepsilon_{\bf k}$ contains $1/S$ corrections  from 
  angle renormalization and Hartree-Fock decoupling of cubic and quartic perturbations, respectively.  
  Ellipses stand for higher-order terms in  the $1/S$ expansion that are neglected in our approximation.  
The three-boson terms are  decay and source vertices 
that are responsible for the anomalous dynamics in the high-field regime.
The dynamical properties of the system are obtained from the interacting magnon 
Green's function, defined as
\begin{equation}
	G^{-1}({\bf k}, \omega)  = \omega - \tilde{\varepsilon}_{\bf k}
	                - \Sigma_{1}({\bf k}, \omega) 
	                - \Sigma_{2}({\bf k}, \omega) 
\label{eq:green}	                
\end{equation}
where $\Sigma_{1, 2}({\bf k}, \omega)$ are the decay and source self-energies  
presented in Fig.~\ref{fig1} and obtained from the second-order treatment of
 Eq.~(\ref{eq:cubic}). Their explicit forms are  
\begin{eqnarray}
	\Sigma_1({\bf k},\omega) & = & \frac{1}{2} \sum_{\bf q}
		\frac{|\Phi_1({\bf k},{\bf q})|^2} {\omega -\varepsilon_{\bf q}-
		\varepsilon_{\bf k-q+Q} + i0} \ ,
	\label{sigma1} \\
	\Sigma_2({\bf k},\omega) & = & -\frac{1}{2} \sum_{\bf q}
		\frac{|\Phi_2({\bf k},{\bf q})|^2} {\omega +\varepsilon_{\bf q}+
		\varepsilon_{\bf k+q-Q} - i0} \ ,
	\label{sigma2}
\end{eqnarray}
with expressions for vertices $\Phi_1$ and $\Phi_2$ given in Appendix~\ref{ap1}. 
In contrast to $H\!=\!0$ case where $\omega$-dependent magnon interactions 
beyond Hartree-Fock have $1/S^2$ smallness, 
they readily occur in $1/S$ order for $H \neq 0$, when the coupling between longitudinal and
transverse modes renders three-boson vertices nonzero.  
Above the threshold field for magnon instability, the self-energy in (\ref{sigma1}) 
also acquires an imaginary component, signifying the occurrence of spontaneous decays.

% --------------------------------------------------------------------
% --------------------------------------------------------------------
\section{Kinematics, singularities and interlayer coupling}
% --------------------------------------------------------------------
% --------------------------------------------------------------------
% --------------------------------------------------------------------
\subsection{Decay boundaries}
% --------------------------------------------------------------------

Although the single- and two-magnon continuum excitations are coupled directly 
by virtue of the cubic terms in Eq.~(\ref{eq:cubic}) at any $H>0$, magnon decays only occur 
above a finite threshold field $H^*$. 
This is due to restrictions provided by the kinematic
conditions, i.e., energy and momentum conservation, that have to be satisfied in each elementary 
decay process
\begin{equation}
	\varepsilon_{\bf k} =\varepsilon_{\bf q}+\varepsilon_{\bf k-q+Q} ,
\label{kinematics}
\end{equation}
where the ordering vector in the momentum conservation is, again, 
due to the staggered canting of spins in the field. 

With the detailed classification of possible 
solutions of Eq.~(\ref{kinematics}) given previously in Refs.~\onlinecite{triangle,Mourigal2010} 
we highlight here two relevant results. First, it can 
be shown that within the Born approximation, that is, neglecting energy renormalization 
of the spectrum $\varepsilon_{\bf k}$ from the linear spin-wave theory result, 
the value of the threshold field 
$H^*$  is independent of the value of the interlayer coupling $\alpha$ and is
the same as in the square lattice case: $H^*\!\approx\!0.76H_s$. 
% --------------------------------------------------------------------
\begin{figure}[t]
\includegraphics[width=0.85\columnwidth]{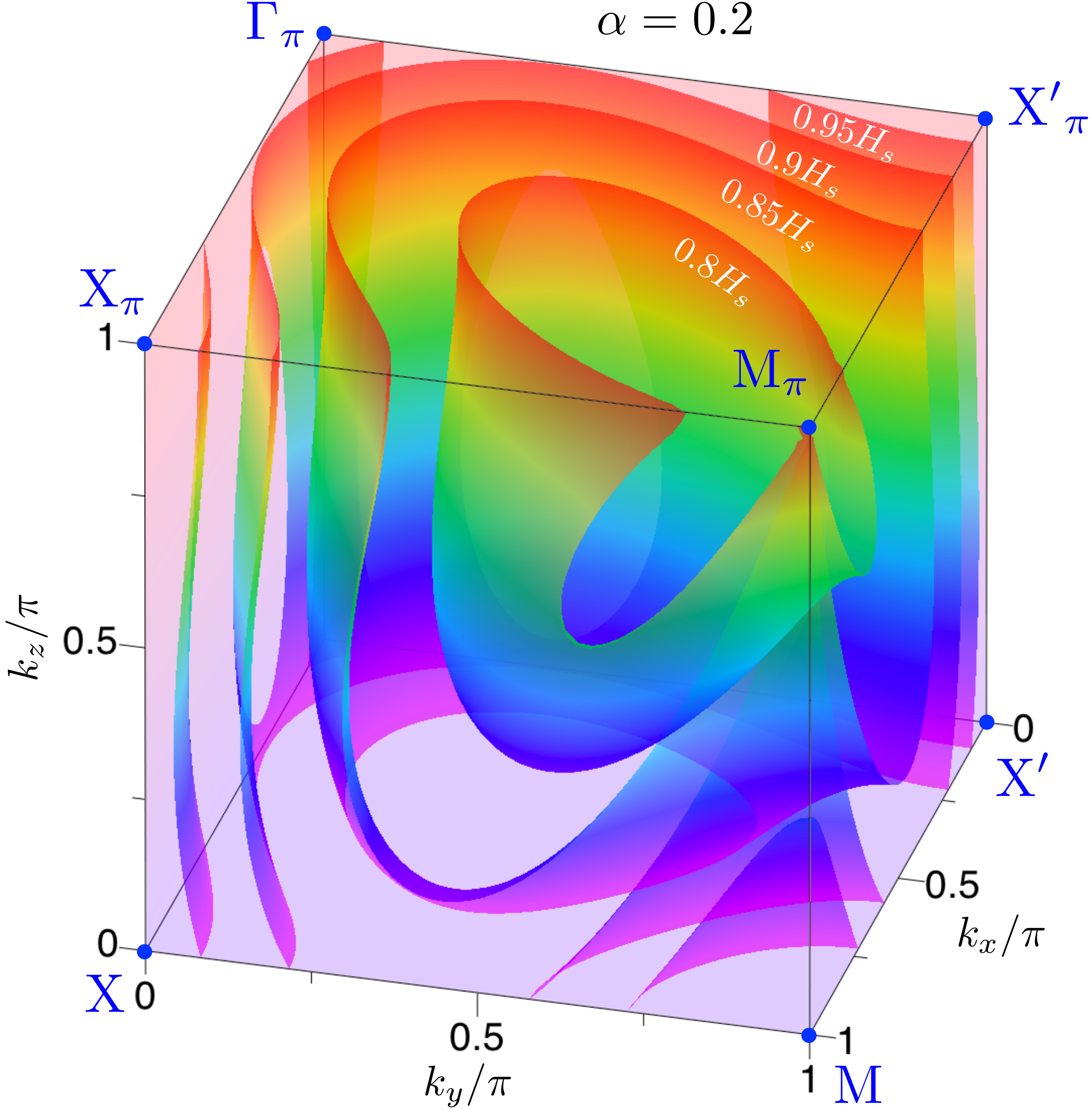}
\caption{(Color online)  Magnon decay regions in one octant of the Brillouin zone for several fields.  
Decays are possible within the portion of the Brillouin zone containing the $\rm{M}_\pi$ point.  
At high fields, decay regions extend beyond the octant and overlap, facilitating larger phase 
space for decays for magnons within the decay region.
}
\label{regions}
\end{figure}
% --------------------------------------------------------------------

Second, for a
given field $H>H^*$, within the same Born approximation that neglects cascade decays, there
exists a sharp boundary in the momentum space separating the region where magnons 
are stable, i.e., cannot satisfy Eq.~(\ref{kinematics}), from the {\it decay region} where they
are unstable towards decays. An example of such {\it decay boundaries} in a 3D setting 
is shown in Fig.~\ref{regions} in one octant of the tetragonal-lattice Brillouin zone for 
$\alpha=0.2$ and several fields.\cite{footnote} 
One can see that the decay region expands with increasing field, covering 
large portions of the Brillouin zone. At high fields, decay regions extend beyond the octant and overlap;
 at $H\agt 0.9H_s$ most magnons are unstable already in the Born approximation.

Taking into account the finite lifetime of the decay products 
in the self-consistent treatment generally leads to blurring away the decay 
boundaries discussed here.\cite{field,Mourigal2010}  
Nevertheless, it is still important to consider them not only because the decays within the 
Born decay region typically remain more intense as they occur in the lower-order process, 
but also because such boundaries correspond to singularities in the spectrum discussed 
next.\cite{footnote1}

% --------------------------------------------------------------------
\subsection{On-shell approximation}
% --------------------------------------------------------------------

\begin{figure}[t]
\includegraphics[width=0.99\columnwidth]{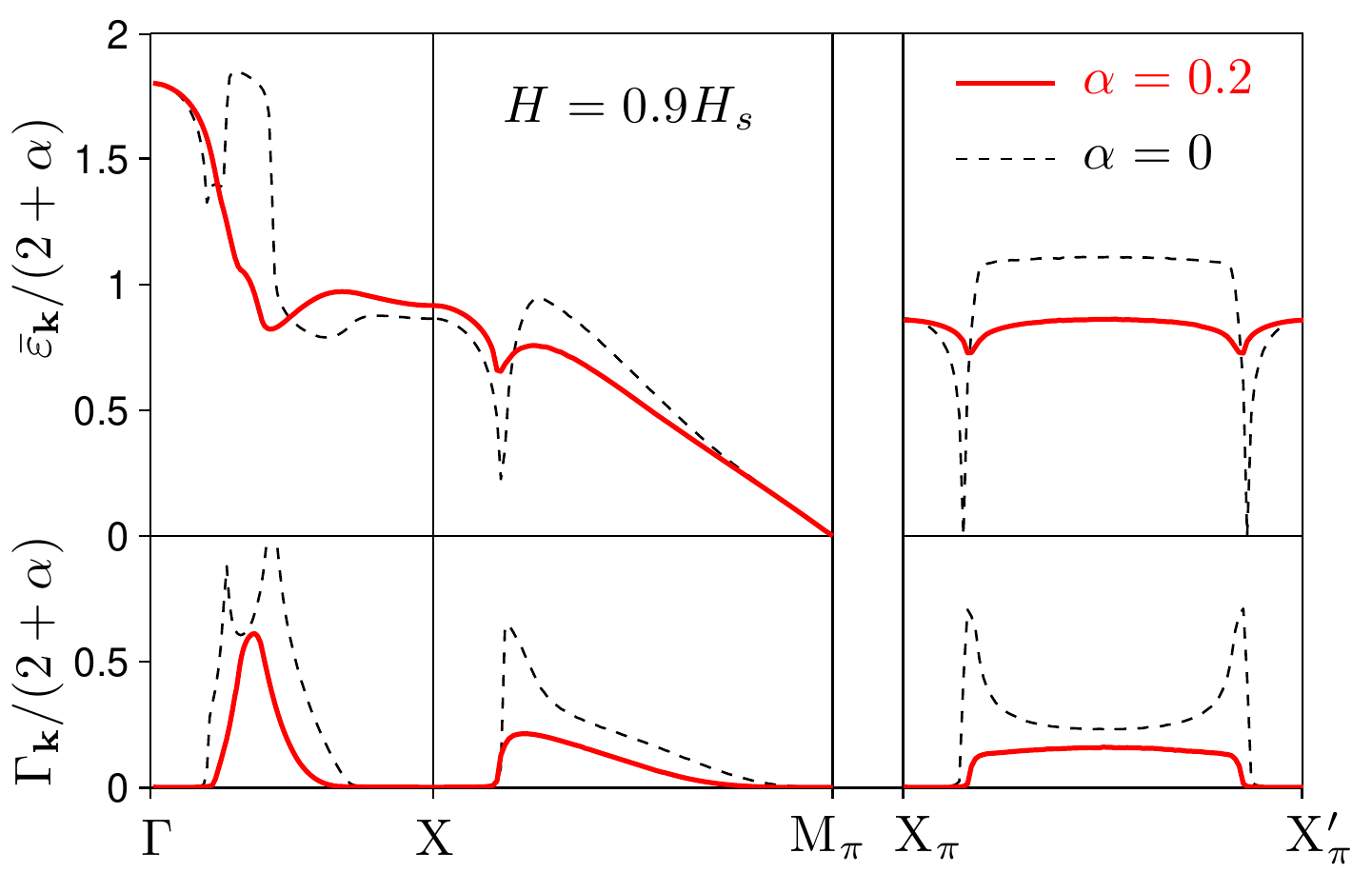}
\caption{(Color online)  Magnon energy and decay rate for $\alpha=0.2$ and for the square-lattice case
($\alpha=0$)
for $H=0.9H_s$ along the selected path in the Brillouin zone (see Fig.~\ref{paths}). 
Upper and lower panels show the renormalized spin-wave energy $\bar{\varepsilon}_{\bf k}$ and 
the decay rate $\Gamma_{\bf k}$ in the on-shell approximation, Eqs.~(\ref{renorm_e}) 
and (\ref{decay}), respectively. Computation was performed with Monte-Carlo numerical integration 
with $5\cdot 10^{7}$ steps.}
\label{onshell}
\end{figure}

Within the standard spin-wave expansion, the {\it on-shell} approximation for the magnon
spectrum consists of setting $\omega=\varepsilon_{\bf k}$ in the self-energies in Eqs.~(\ref{sigma1}), 
(\ref{sigma2}), leading to the renormalized spectrum
\begin{equation}
\bar{\varepsilon}_{\bf k}=\tilde{\varepsilon}_{\bf k}+
{\rm Re}\,\Sigma_1({\bf k},\varepsilon_{\bf k})+\Sigma_2({\bf k},\varepsilon_{\bf k}),
\label{renorm_e}
\end{equation}
and the decay rate
\begin{equation}
	\Gamma_{\bf k}=\frac{\pi}{2}\sum_{\mathbf{q}} |\Phi_{1}(\mathbf{k,q)}|^2 
	               \delta\bigl(  \varepsilon_{\bf k} -\varepsilon_{\bf q}
	                            -\varepsilon_{\bf k-q+Q} \bigr).
\label{decay}
\end{equation}
We note that this approach implicitly suggests that the dynamical response can be 
well approximated by the quasiparticle
peaks at renormalized energies $\bar{\varepsilon}_{\bf k}$ with lorentzian broadening defined by
$\Gamma_{\bf k}$. As was demonstrated in Ref.~\onlinecite{field,Mourigal2010} for the 2D square-lattice 
and in Ref.~\onlinecite{triangle} for the triangular-lattice
cases, this approach, which is normally very accurate even quantitatively, breaks down rather
dramatically, showing divergences in $\bar{\varepsilon}_{\bf k}$ or
$\Gamma_{\bf k}$ for the momenta belonging to various contours in ${\bf k}$-space.
We reproduce some of them in our Fig.~\ref{onshell}, shown by dashed lines.
These singularities were understood as coming from the van Hove singularities in the two-magnon
continuum through coupling with the single-magnon branch. 
One can see in the expression for $\Gamma_{\bf k}$ in Eq.~(\ref{decay}) that 
once the continuum and the single-magnon branch overlap in some portion of the Brillouin zone,
$\varepsilon_{\bf k}$ is effectively exploring the two-magnon density of states as a function of ${\bf k}$. 
If a singularity in the latter is met, it will be reflected in $\Gamma_{\bf k}$ and, by the 
Kramers-Kronig relation, in $\bar{\varepsilon}_{\bf k}$. 

Generally, there are two types of singularities that can occur, associated with the minima (maxima) 
and saddle points of the two-magnon continuum, respectively. The decay boundaries discussed
previously are {\it necessarily} the surfaces
% (contours in 2D) 
of such singularities 
because they must correspond to the intersection with a minimum of the two-magnon continuum, 
forming the locus of points where decay conditions are first met. 
 These {\it decay threshold} 
singularities correspond to the step-like behavior in $\Gamma_{\bf k}$
and the logarithmic divergence in $\bar{\varepsilon}_{\bf k}$.\cite{field,triangle,Mourigal2010} 
For the ${\bf k}$-points already inside
the decay region, there is a possibility of meeting a saddle point of the continuum, in which
case the real part of the spectrum $\bar{\varepsilon}_{\bf k}$ 
experiences a jump and $\Gamma_{\bf k}$ a 
logarithmic singularity, see the 2D data in Fig.~\ref{onshell}.
%------------------------------------------------------------------------------
% -------------------------------------------------------------------------------

Essential for the present study is the change in the behavior of these singularities for a  moderate
interlayer coupling. One can expect from the density of states argument that both types
of singularities must become weaker and change from logarithmic/step-like to 
square-root-like,\cite{triangle} 
resulting in softening of the singular behavior of $\Gamma_{\bf k}$ and $\bar{\varepsilon}_{\bf k}$. 
This effect is demonstrated in Fig.~\ref{onshell} for several directions in the 3D Brillouin zone
compared with equivalent cuts for the square-lattice case.
The full account of analytical and numerical details of the on-shell calculations 
is given in Appendix A. 
The definitions concerning the Brillouin zone path in Fig.~\ref{onshell}  can be found in Fig. 7.
It is rather remarkable that the singularity of the saddle-point type is completely wiped out by the 
moderate interplane coupling, see panel $\Gamma-\rm{X}$  in Fig.~\ref{onshell}.
 While the decay threshold singularities associated with the boundaries in Fig.~\ref{regions},  
are not fully alleviated, they are considerably diminished.   
For both types of singularities, further increase of the interplane coupling 
%from the one shown in Fig.~\ref{onshell} 
leads to 
gradual changes in the on-shell spectrum and decay rates;  singularities
shown in  Fig.~\ref{onshell} for $\alpha=0.2$ remain largely unchanged.  

While singularities are diminished, the magnon damping 
remains considerable. Since both the decay boundaries and $\Gamma_{\bf k}$ are obtained
within the Born approximation, the decay rate in Fig.~\ref{onshell} is non-zero within the 
corresponding decay region
outlined in  Fig.~\ref{regions} for $H=0.9H_s$, demonstrating that magnons are
unstable in most of the Brillouin zone, while the Goldstone mode at $M_\pi$ $[{\bf k}\!=\!(\pi,\pi,\pi)]$ 
and the uniform precession mode at $\Gamma$ $[{\bf k}\!=\!(0,0,0)]$ remain well-defined. 

Previous works on regularization of singularities in the spin-wave spectrum in the 2D square-lattice
case\cite{field,Mourigal2010} 
were based on partial dressings of magnon Green's functions in the 
one-loop diagrams of Fig.~\ref{fig1}. Generally, the difficulty of a consistent implementation
of such schemes is the opening of an unphysical gap in the acoustic branch. The latter problem 
was avoided in Ref.~\onlinecite{Mourigal2010} by performing self-consistency only in the 
imaginary part of the magnon self-energy, applicable for larger $S\!\geq\! 1$ where 
the real part of renormalization can be deemed small. The downside of this approach is that it remains 
essentially on-shell, enforcing Lorentzian shapes of the quasiparticle peaks.
In a sense, what is being argued by our Fig.~\ref{onshell} is that 
introducing finite interplanar coupling represents an alternative to the self-consistency schemes in 
regularizing singularities, without restrictions on the spectral shapes and at a fraction of 
the computational cost.

% --------------------------------------------------------------------
\subsection{Spectral function}
% --------------------------------------------------------------------

The one-loop self-energies $\Sigma_{1, 2}({\bf k}, \omega)$ 
in Eqs.~(\ref{sigma1}) and (\ref{sigma2}) originate from the coupling to the two-magnon 
continuum. Because of their $\omega$-dependence,  one can  obtain  significantly richer
dynamical information, beyond the quasiparticle pole-like state 
expected from the  on-shell approach above, by considering the diagonal component of the 
spectral function,
\begin{equation}
{A}({\bf k}, \omega) = -\frac{1}{\pi}\,{\rm Im}\,G({\bf k}, \omega),
\label{eq:spectral}
\end{equation}
where $G({\bf k}, \omega)$ is the interacting magnon Green's function defined in Eq.~(\ref{eq:green}). 
Because of the interactions, the spectral function is expected to exhibit an incoherent 
component which should reflect the two-magnon continuum states in addition to
a quasiparticle peak that may be broadened.
The spectral function is directly related to the 
dynamical structure factor $S({\bf k},\omega)$ measured in neutron-scattering experiments, 
which we discuss in detail in the next Section.
Strictly speaking, the consideration of $A({\bf k},\omega)$ goes beyond the $1/S$ expansion as 
$\omega\neq\varepsilon_{\bf k}$, yet it is free from the complications due to higher-order diagrams
mentioned above. 

We should note that in the pure 2D, $S=1/2$ case,  non-selfconsistent 
calculation of $A({\bf k},\omega)$ does exhibit incoherent subbands,\cite{field} 
but suffers from the high density of states of the two-magnon continuum 
associated with the threshold singularities considered above. Because of level repulsion, the 
renormalized quasiparticle peaks escape the unrenormalized two-magnon continuum 
and preclude an accurate determination of the dynamical structure factor, except for fields in close
vicinity of the saturation field where the renormalization becomes weaker.
However, this is not the case in the triangular-lattice antiferromagnet in zero field,\cite{triangle} 
where the thresholds are controlled by the emission of the Goldstone magnons associated  
with much smaller density of states.

Similar to the weakening of the threshold singularities in 3D, one can expect 
weaker repulsion between the single-magnon branch and the minima of the two-magnon continuum.
As will be shown in the next Section, while the quasiparticle energy shift and corresponding 
escape from the continuum are still present to some degree, they are considerably smaller 
than in the 2D case. Therefore, once again, introducing finite interplanar coupling 
provides an alternative to the self-consistency schemes, allowing for a detailed analysis 
of the effects of broadening and weight redistribution in the spectrum.

% --------------------------------------------------------------------
% --------------------------------------------------------------------
\section{Dynamical Structure Factor}
% --------------------------------------------------------------------
% --------------------------------------------------------------------
% -------------------------------------------------------------------------
\begin{figure*}[t]
\includegraphics[width=1.8\columnwidth]{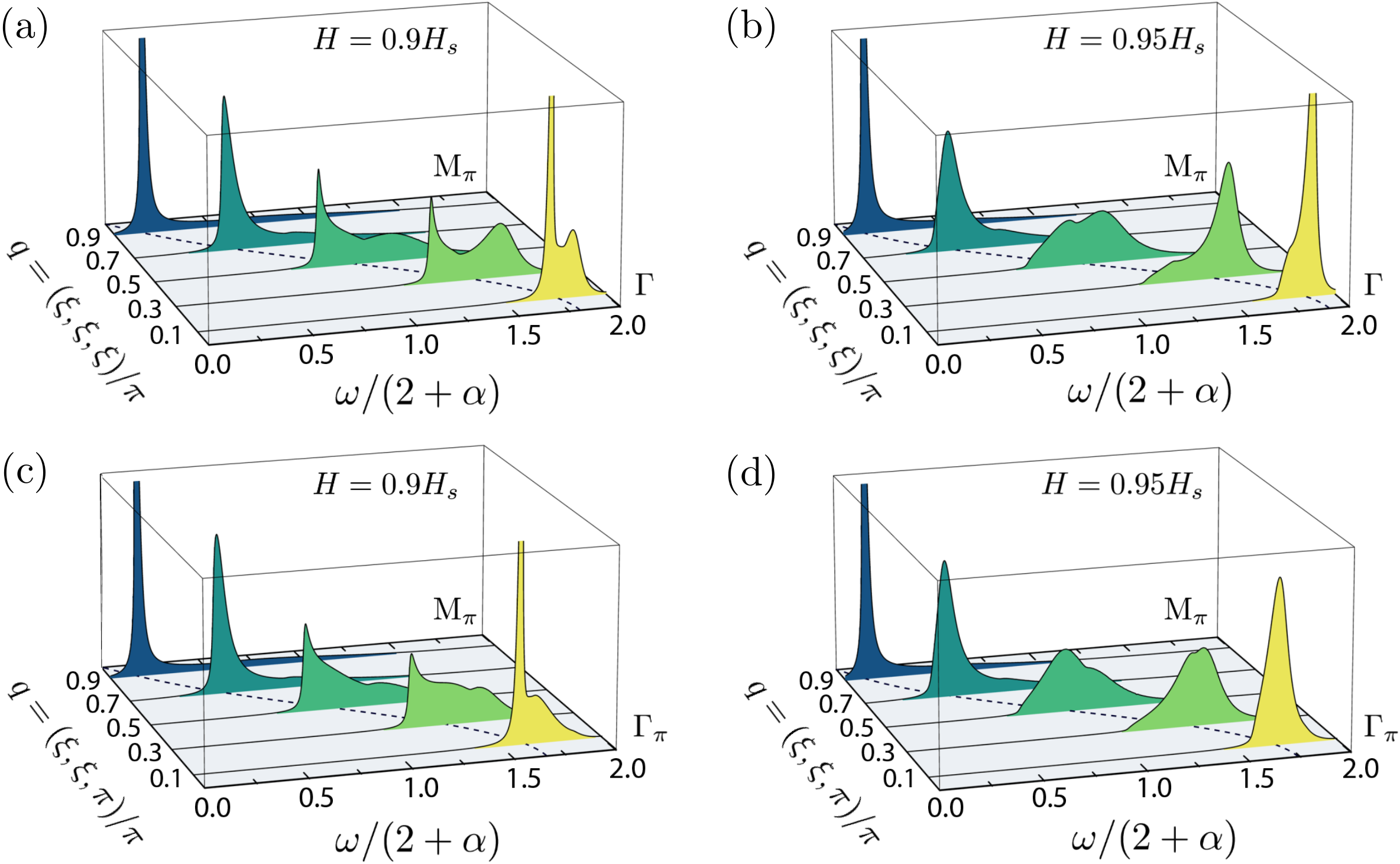}
\caption{(Color online) Profiles of the perpendicular component of the dynamical 
structure factor $S_\perp ({\bf k}, \omega)$, Eq.~(\ref{Sperp}), vs $\omega$ 
for selected ${\bf k}$-points along the main diagonal of the Brillouin zone 
(from $\Gamma$ to ${\rm M}_\pi$), panels (a) and (b), and
 along the diagonal in the $k_z =\pi$ plane (from $\Gamma_\pi$ to ${\rm M}_\pi$), 
 (c) and (d), for fields $0.9H_s$ and $0.95H_s$, and $S=1/2$. For high-symmetry ${\bf k}$-points
 notations, see Figs.~\ref{regions} and \ref{paths}.
 Dashed lines indicates the linear spin-wave dispersion $\varepsilon_{\bf k}$ along the same paths.  
}
\label{pro}
\end{figure*}
% -------------------------------------------------------------------------

The inelastic neutron-scattering cross section is proportional to a linear combination 
of the components of the spin-spin dynamical correlation function\cite{Squires}
\begin{equation}
S^{\alpha\beta}({\bf k},\omega) =
\int_{-\infty}^{\infty}  \frac{dt}{2\pi}\,
\left\langle S^{\alpha}_{\bf k }(t) S^{\beta}_{-\bf k } \right\rangle\, e^{i\omega t}
\label{Salphabeta}
\end{equation}
where $\alpha,\beta=\{x_0,y_0,z_0\}$ span the Cartesian directions of the laboratory frame. 
For simplicity,  in the following discussion we ignore
additional momentum-dependent polarization factors from experimental details, on which 
coefficients of such linear combination can depend on, and show our results for what we will refer to as 
the ``full''  dynamical structure factor
\begin{equation}
	S({\bf k},\omega) = S^{x_0x_0}({\bf k},\omega) +S^{y_0y_0}({\bf k},\omega)+S^{z_0z_0}({\bf k},\omega),
\label{Sfull}
\end{equation}
as well as its component perpendicular to the external field, 
the latter is directed along the $z_0$-axis as before, 
\begin{equation}
	S_\perp({\bf k},\omega) = S^{x_0x_0}({\bf k},\omega) +S^{y_0y_0}({\bf k},\omega).
\label{Sperp}
\end{equation}
The choice of these particular examples is both illustrative and general, as it exposes 
all the essential components of $S({\bf k},\omega)$.

The dynamical structure factor is naturally written in a laboratory reference frame while the magnon 
operators were introduced in the rotating frame that aligns local spin quantization axis on 
each site in the direction given by a classical configuration of spins canted in a field.
Thus, the relation of the magnon spectral function, $A({\bf k},\omega)$ 
of Eq.~(\ref{eq:spectral}), to $S({\bf k},\omega)$ 
for canted spin structures is via a two-step transformation: rotation of spins from laboratory to local
frame and bosonization of spin operators.\cite{Mourigal2010}
In the lowest $1/S$-order, this procedure becomes rather straightforward since most complications
such as off-diagonal terms in the spin Green's functions  
 can be justifiably dropped as they only contribute in the
order $1/S^2$ or higher. With that, performing the second transformation first, 
components of the dynamical structure factor 
in the {\it local} quantization frame are readily given by\cite{Mourigal2010}
\begin{eqnarray}
	S^{xx}({\bf k},\omega) & \approx &    \nonumber
     	\pi S \Lambda_+ \: (u_{\bf k } + v_{\bf k })^2 A ({\bf k},\omega), \\
	S^{yy}({\bf k},\omega) & \approx & 
     	\pi S \Lambda_- \: (u_{\bf k } - v_{\bf k })^2 A ({\bf k},\omega), 
\label{eq:dsf}     	\\
     S^{zz}({\bf k},\omega) & \approx &     \nonumber
     	\pi\sum_{\bf q} (u_{\bf q} v_{\bf k-q} + v_{\bf q} u_{\bf k-q})^2   \delta(\omega-
     	\varepsilon_{\bf q}-\varepsilon_{\bf k-q}), \nonumber 
\end{eqnarray}
where $x,y,z$ now refer to local axes,  $u_{\bf k}$ and 
$v_{\bf k}$ are parameters of the Bogoliubov transformation, and 
$\Lambda_{\pm} \!=\! (1 \!-\! (2n\!\pm\!\delta)/2S)$ contain the $1/S$ corrections from the
Hartree-Fock spin reduction factors, see Appendix A. 

Strictly speaking, in these expressions for the diagonal terms of $S^{\alpha\alpha}$ 
we have kept contributions in excess of the leading order of $1/S$ expansion,
as both the  Hartree-Fock corrections to $S^{xx}$ and $S^{yy}$ as well as the $S^{zz}$ term 
itself are  of higher order. In addition to being easy to include them into our consideration, they also 
offer us an opportunity to demonstrate their relative unimportance. Thus, 
 the longitudinal component $(S^{zz})$ in the given order contains a ``direct'' contribution of
 the two-magnon continuum to the structure factor. Apart from being broadly distributed over the
  $({\bf k},\omega)$ space compared to a more sharply concentrated $A({\bf k},\omega)$, 
its contribution to the scattering turned out to be diminishingly small compared to the two transverse components $S^{xx(yy)}$ in the considered high-field regime. Besides being an effect
of the higher $1/S$ order, this is also due to its dependence on
$v^2$ ($\propto\! \cos^4 \theta\!\ll\!1$), where $\theta$ is the spins' canting angle which tends
to $\pi/2$  as the saturation field is approached. 

Completing the connection of the local  form of correlation functions to the laboratory 
ones gives the final relation of  $A({\bf k},\omega)$ to $S({\bf k},\omega)$ 
 \begin{eqnarray}
S^{x_0x_0}({\bf k},\omega)  & \approx &  \sin^2\!\theta \,
S^{xx}({\bf k},\omega) +  \cos^2\!\theta\,
S^{zz}({\bf k}-{\bf Q},\omega)\,,
\nonumber \\
S^{z_0z_0}({\bf k},\omega)  & \approx &
\cos^2\!\theta\, S^{xx}({\bf k}-{\bf Q},\omega) +
\sin^2\!\theta \, S^{zz}({\bf k},\omega)\,,
\nonumber               \\
S^{y_0y_0} ({\bf k},\omega) & = & S^{yy}({\bf k},\omega) \ ,
\label{SQW1}	
\end{eqnarray}
where the cross-terms in spin-spin correlation function 
($S^{xz}$ and $S^{zx}$) in the r.h.s. were omitted under the same premise of 
not contributing in the lowest order.
 These last expressions demonstrate that due to external field, 
 spin-canting redistributes the spectral weight over two transverse modes,\cite{Birgeneau,Mourigal2010} 
 referred to as the ``in-plane'' and the ``out-of-plane'' modes, 
 corresponding to the momentum ${\bf k}$ and to the momentum 
 ${\bf k}-{\bf Q}$ (and to fluctuations in and out of the plane
 perpendicular to applied field), respectively.  
 As can be expected form the pre-factor $\cos^2\theta$ in Eq.~(\ref{SQW1}), 
 strong magnetic field suppresses the out-of-plane mode until it disappear entirely at $H_s$. 
 In the $S({\bf k},\omega)$ plots that will follow, the out-of-plane contribution appears as a
 ``shadow'' of the main signal, shifted by the ordering vector.

From Eqs.~(\ref{SQW1}) and (\ref{eq:dsf}) one can see that our choice of the perpendicular 
component of the structure factor $S_\perp({\bf k},\omega)$ in Eq.~(\ref{Sperp}) as a representative  
example becomes particularly simple. Since it is not contaminated 
by the ${\bf k}-{\bf Q}$ shadow component and the contribution of the 
longitudinal $S^{zz}({\bf k},\omega)$ to it is exceedingly small in the considered 
field range, it is closely approximated by
\begin{equation}
S_\perp({\bf k},\omega) \approx  \pi S 
f_{\bf k }  A ({\bf k},\omega),
\label{Sperp1}
\end{equation}
where $f_{\bf k}\!=\!\left[\sin^2\!\theta\Lambda_+ \: (u_{\bf k } \!+ \!v_{\bf k })^2\!+\!
\Lambda_- \: (u_{\bf k } \!- \!v_{\bf k })^2\right]$ is the ${\bf k}$- and field-dependent intensity factor.

Our Fig.~\ref{pro} presents the perpendicular component of the dynamical structure factor 
$S_\perp({\bf k},\omega)$ for $S\!=\!1/2$, fields of $0.9H_s$ and $0.95H_s$, 
and for several ${\bf k}$-points along two representative directions: main diagonal of the 
Brillouin zone (from $\Gamma$ to ${\rm M}_\pi$), Fig.~\ref{pro}(a) and (b), and 
diagonal of  the $k_z=\pi$ plane (from $\Gamma_\pi$ to ${\rm M}_\pi$), Fig.~\ref{pro}(c) and (d), respectively. 
To obtain these profiles,  integrals in the self-energies in Eqs.~(\ref{sigma1}) and (\ref{sigma2}) 
and in $S^{zz}({\bf k},\omega)$ in Eq.~(\ref{eq:dsf}) were 
computed using Mathematica via adaptive quasi-Monte-Carlo
without symbolic preprocessing, over $5\cdot 10^6$ points,
maximum recursion of $10^4$, and  an accuracy goal of 4 digits.   
Each presented ${\bf k}$-cut contains 400 points in $\omega$ and an artificial broadening 
$\delta/(2\!+\!\alpha)=10^{-2}J$ was used. As we discussed, contribution of 
the longitudinal term $S^{zz}$ is not visible on the scale of the plots. 
Dashed lines show  linear spin-wave dispersions for the given paths.
% -------------------------------------------------------------------------
\begin{figure*}[t]
\includegraphics[width=1.8\columnwidth]{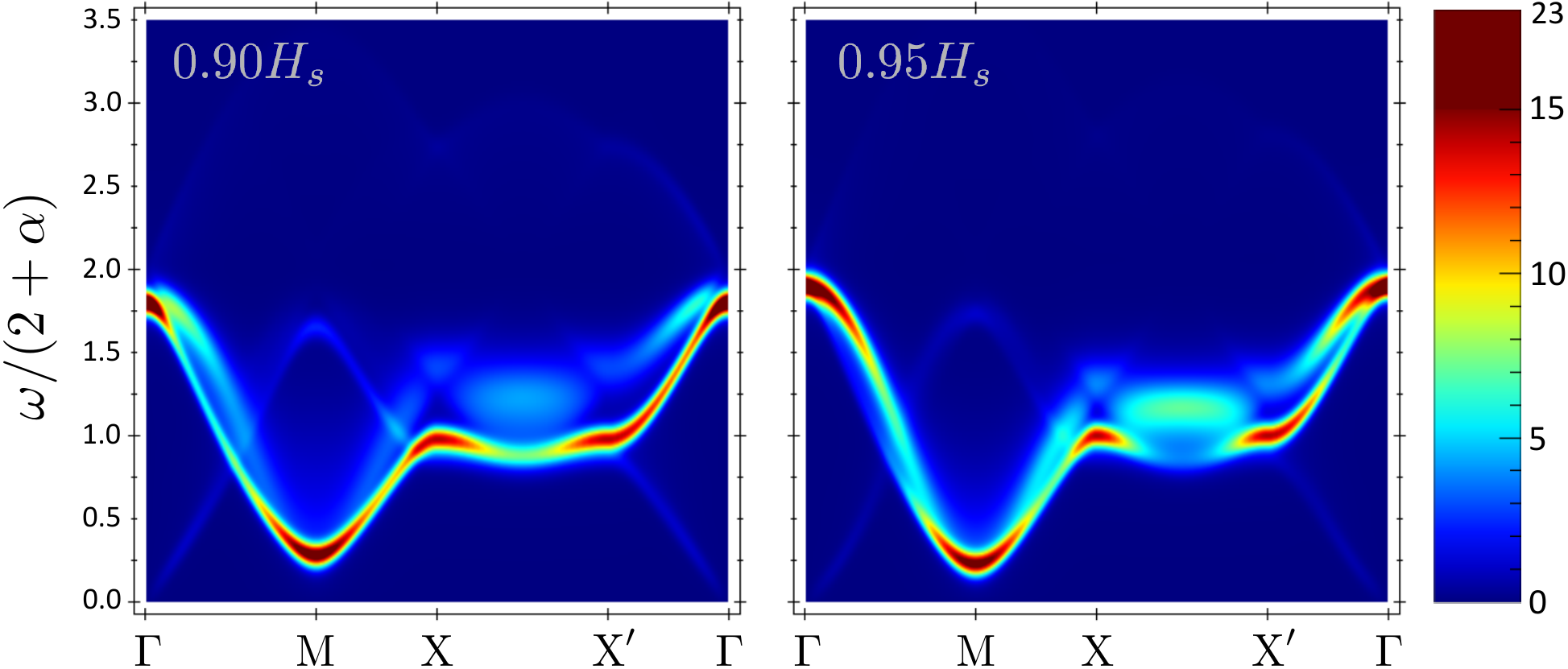}
\caption{(Color online)
Intensity plots of the dynamical structure factor $S({\bf k},\omega)$, Eq.~(\ref{Sfull}),  
for fields $0.9H_s$ and $0.95H_s$, and $S=1/2$,
for the ${\bf k}$-path in the $k_z\!=\!0$ plane, see Fig.~\ref{paths}.  
Gaussian convolution was performed in $\omega$ 
with a $\sigma$-width of $\delta \omega\!=\!0.1 J$. 
}
\label{skw1}
\end{figure*}
% -------------------------------------------------------------------------

These $S_\perp({\bf k},\omega)$ profiles demonstrate that the spectrum broadening due to magnon
 decays in the high-field regime is found abundantly throughout the Brillouin zone. 
In addition, clear double-peak structures, similar to that seen in the self-consistent spin-wave 
calculations\cite{field} and emphasized in the QMC study\cite{Olav08} 
for the 2D square-lattice case, occur along the main diagonal, Fig.~\ref{pro}(a), (b). 
At $H\!=\!0.90 H_s$, some sharp magnon peaks remain in the vicinity of the $\Gamma$-point,
in agreement with the decay boundaries in Fig.~\ref{regions}. Peaks also get narrower
at the approach of the Goldstone ${\rm M}_\pi$ point. 
Although there seems to be some sharpness in the structure of $S_\perp({\bf k},\omega)$ data for 
$H\!=\!0.90 H_s$ in the middle 
of the diagonals, these are not associated with quasiparticle peaks, but with the remaining singular 
behavior of the one-loop self-energies. 

A noteworthy feature of the shown results is a  strong 
field-evolution of $S_\perp({\bf k},\omega)$, demonstrating a rather dramatic redistribution 
of spectral weight between the broadened single-particle and incoherent part of the spectrum, related 
to the two-magnon continuum.  Another important observation is the relative smallness of the energy 
shift of the spectral-function leading edge in $\omega$ compared to the linear spin-wave energy
 $\varepsilon_{\bf k}$, shown by the dashed lines. This, together with a close similarity of the
shown non-selfconsistent results to the self-consistent ones in the 2D square-lattice case,\cite{field} 
is an indication of the reliability of current approach, thus supporting our expectation that  
presented results should serve as a fair representation of the realistic structure factor.

\begin{figure*}[t]
\includegraphics[width=1.8\columnwidth]{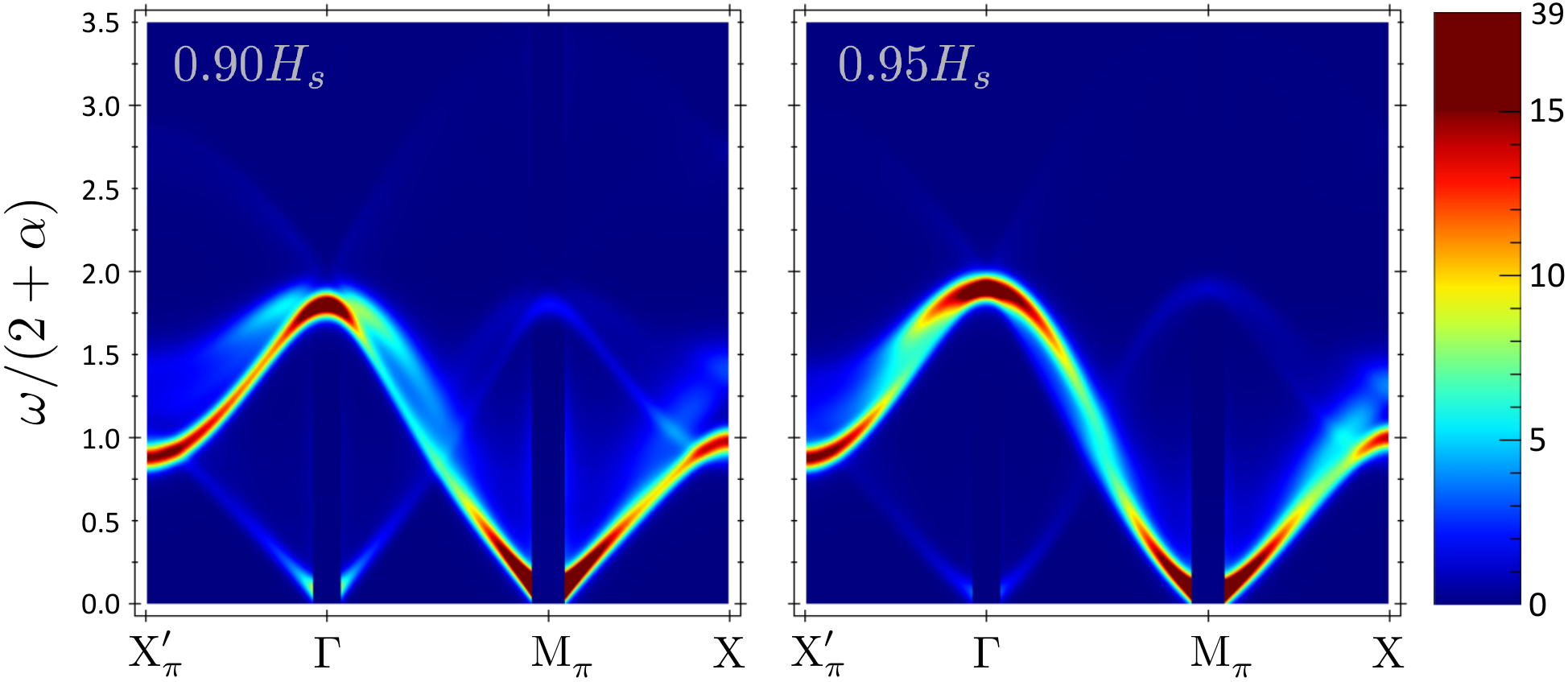}
\caption{(Color online) Same as in Fig.~\ref{skw1}, for a different path, see Fig.~\ref{paths}. 
The region near $\rm{M}_\pi$ in ${\bf k}$ and ${\bf k}-{\bf Q}$ modes are cut due 
to spurious divergences, see text.  
Note the higher cutoff in intensity due to a  formfactor divergent at
$\rm{M}_\pi$-point.  
}
\label{skw2}
\end{figure*}

\begin{figure}[b]
\includegraphics[width=0.7\columnwidth]{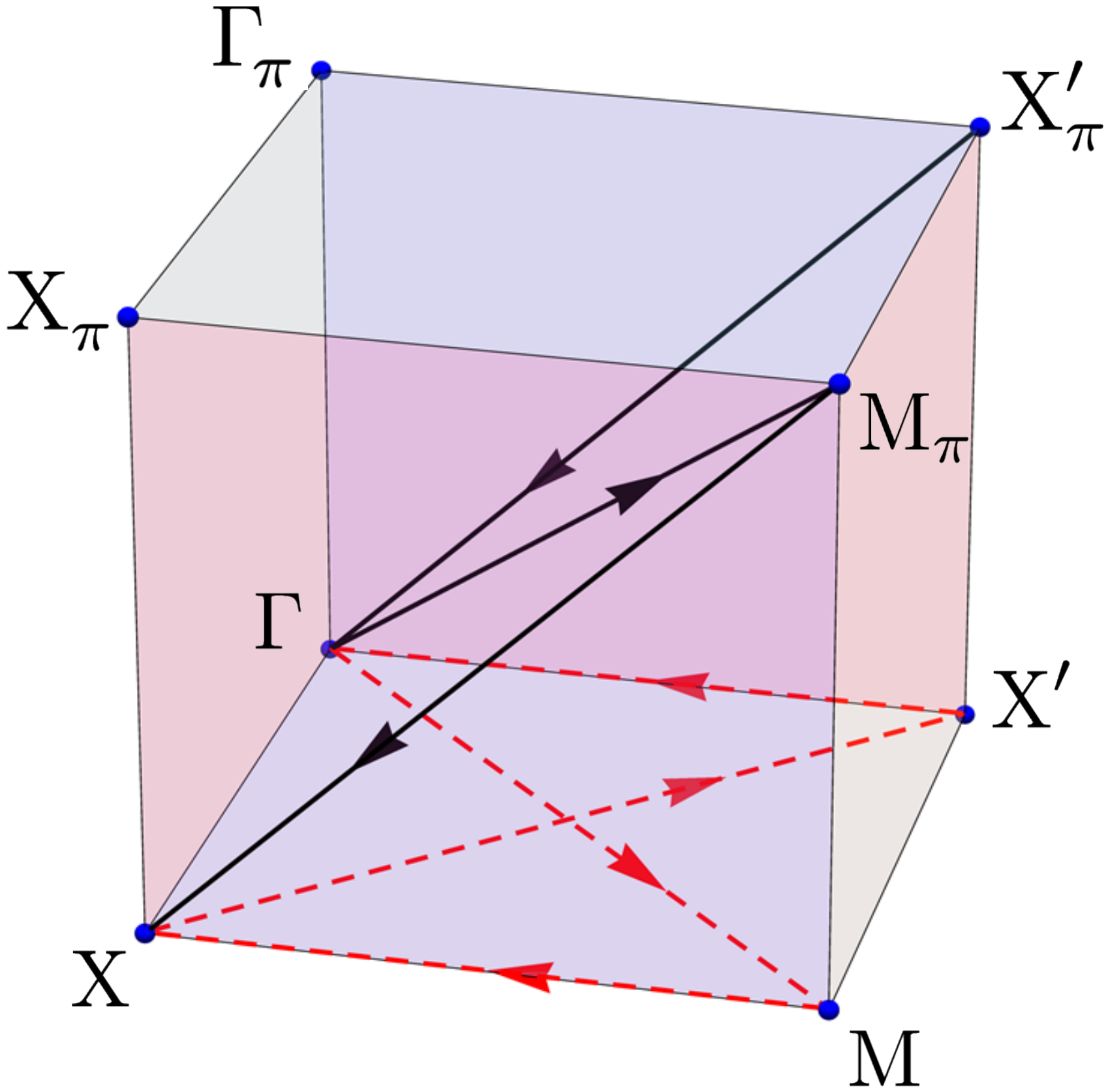}
\caption{(Color online) Brillouin zone octant showing the ${\bf k}$-paths 
traversed in Figs.~\ref{skw1} and \ref{skw2}.  Path in Fig.~\ref{skw1} lies in the $k_z = 0$ plane 
and goes along the dotted line: 
$\rm{\Gamma} \rightarrow \rm{M} \rightarrow \rm{X} \rightarrow \rm{X'} \rightarrow\Gamma$.  
Path in Fig.~\ref{skw2}  goes between the $k_z = 0$ and $k_z = \pi$ planes along the zig-zag solid line: 
$\rm{X'}_{\pi} \rightarrow \rm{\Gamma} \rightarrow \rm{M}_{\pi} \rightarrow \rm{X}$.  
}
\label{paths}
\end{figure}

In Figs.~\ref{skw1} and \ref{skw2}, intensity plots of the 
full dynamical structure factor, $S({\bf k},\omega)$ in
Eq.~(\ref{Sfull}), for $S\!=\!1/2$, fields of $0.9H_s$ and $0.95H_s$, 
and for several  representative paths in the Brillouin zone are shown.
In Fig.~\ref{skw1},  the ${\bf k}$-path is entirely in the $k_z = 0$ plane and in Fig.~\ref{skw2} 
the momentum  traces an inter-layer path, both shown in Fig.~\ref{paths}. 
As we have discussed above,  in addition to the main features demonstrated in Fig.~\ref{pro} for 
$S_\perp({\bf k},\omega)$ that are directly related to the magnon spectral
function $A({\bf k},\omega)$, the out-of-plane transverse mode contribution is visible as 
a shadow, shifted by the ordering vector.
The longitudinal ($S^{zz}$) contributions are hardly noticeable on the scale of the plots as before. 
In Fig.~\ref{skw2}, proximity to the Goldstone mode at $M_\pi$
causes unphysical divergences, amplified by the divergent intensity factor $f_{\bf k}$ in Eq.~(\ref{Sperp1}),
that necessitate removal of this region in the plot.

Integrals in the self-energies for these plots were computed by the same method as in Fig.~\ref{pro}, 
over a maxium number of points of $2\cdot 10^5$, 
with the same accuracy goal of 4 digits.  There are 150 points along each segment in the ${\bf k}$-path,  
each with 350 points in $\omega$.  All integrations were performed in parallel over 8 logical threads. 
The same upper cut-off in intensity has been used in both plots with additional red color fill up to 
the extent of the highest peak, whose value is shown in the color sidebars.
Additionally, we have performed
Gaussian convolution in the $\omega$-direction with a $\sigma$-width of $\delta \omega\!=\!0.1 J$. 
This step is intended to mimic the effect of a realistic experimental resolution, from which 
we can also draw quantitative predictions of the relative strength of the quasiparticle and 
incoherent part of the spectrum. 

Despite the provided broadening by the finite energy resolution, complex spectral lineshapes, 
very much distinct from the conventional quasiparticle peaks,  are clearly visible in 
Figs.~\ref{skw1} and \ref{skw2}. This demonstrates that the 
effects of spectral weight redistribution and broadening due to spontaneous decays are substantial
and should be readily observed in experiment.  

Since $\Gamma$, $\rm{M}$, $\rm{X}$, and $\rm{X}'$ points in Fig.~\ref{skw1} and 
$\Gamma$, $\rm{X}$, and $\rm{X}'_\pi$ 
points in Fig.~\ref{skw2}  are outside of the Born approximation decay regions according to 
Fig.~\ref{regions}, spectrum in their vicinities exhibits well-defined quasiparticle peaks, 
broadened by our ``instrumental'' resolution.  However, away from them,  the structure factor
demonstrates a variety of unusual features, including the already discussed double-peak lineshape
for $\Gamma \rm{M}$ direction in Fig.~\ref{skw1} and $\Gamma \rm{M}_\pi$ direction in Fig.~\ref{skw2}, the
latter path also shown in $S_\perp({\bf k},\omega)$ previously.

Along some of the ${\bf k}$-directions for $H=0.9H_s$ in both Fig.~\ref{skw1} and ~\ref{skw2}, 
the renormalized quasiparticle peaks escape the unrenormalized two-magnon continuum and survive 
despite being formally within the decay region.  Yet,  these peaks are accompanied 
by continuum-like subbands that accumulate significant weight.  
Upon increase of the field, 
the continuum overtakes the single-particle branches in the large regions of the Brillouin zone, 
washing them out and creating more double-peak structures, e.g.,  very distinctly 
along the $\rm{X}'\Gamma$ (Fig.~\ref{skw1}) and $\rm{X}'_\pi\Gamma$ (Fig.~\ref{skw2}) directions. There is
a particularly spectacular advance of the continuum along the $\rm{XX}'$ line in Fig.~\ref{skw1}, where 
magnon broadening  is most intense with dramatic spectral weight transfer to high-energies.  

% -------------------------------------------------------------------------
% -------------------------------------------------------------------------
% ---------------------------------------------------------------------------
% ----------------------------------------------------------------------------

We note that the regions where decays are present in the on-shell solutions discussed
in Sec. III  often correspond closely to the ``washout'' regions of the intensity plots, 
although the latter draw a much richer picture by exposing a complex $\omega$-structure. 
The remaining 3D singularities in the real part of the on-shell energy, exemplified in  Fig.~\ref{onshell},
indicate where the two-magnon continuum overlaps with the single-magnon branch, also 
in a qualitative agreement with the intensity plots, thus providing a complimentary information. 
Previous examination of the 2D square-lattice case in Ref.~\onlinecite{Mourigal2010} 
showed broadening patterns similar to the ones in Figs.~\ref{skw1} and Fig.~\ref{skw2}
for corresponding paths of the Brillouin zone, although 
the $\omega$-structure differs significantly as the self-consistent approach of that work 
has enforced the Lorentzian shapes of the spectral function.  

The one-loop approximation used in this work takes into account the real part of the 
self-energy corrections, allowing the renormalized single-magnon branch 
to escape some of the decay region. Therefore, the obtained dynamical structure factor 
is likely to overestimate the field values needed to reach a significant overlap between 
the two-magnon continuum and the single magnon mode throughout the Brillouin zone. 

Altogether, examination of the dynamical structure factor reveals rich features consisting of 
dramatic redistributions of magnon spectral weight. This ranges from the appearance of 
double-peak lineshapes to a full suppression of quasiparticle peaks throughout large portions 
of the Brillouin zone. These features fully survive convolution with moderate 
energy resolution and should therefore be accessible to state-of-the-art neutron scattering experiments.

\section{Conclusions}

In this work, we have 
provided a theoretical consideration of high-field dynamics in the 
quasi-2D tetragonal $S\!=\!1/2$ Heisenberg antiferromagnet with interlayer coupling corresponding 
to realistic materials. We have presented quantitative predictions for the dynamical structure 
factor for a representative interlayer coupling ratio $J'/J\!=\!0.2$ 
along several  paths in the Brillouin zone. We have demonstrated 
within the framework of spin-wave theory that the finite 3D interlayer coupling, present to 
some degree in all realistic materials, largely 
mitigates singularities of high-field corrections that are complicating similar analysis in the
purely 2D case and necessitate a self-consistent regularization in the latter. 

We argue that introducing finite interplanar coupling effectively
provides an alternative to the self-consistency schemes, 
allowing for a detailed analysis of the effects of broadening and weight redistribution in the spectrum
without the use of self-consistency. 
This relatively simple procedure requires considerably less computation than 
self-consistent calculations, follows the spirit of the $1/S$ spin-wave expansion 
and provides a physical representation of the excitation spectrum. 
Our results therefore consist in a detailed picture of the latter and offer 
clear predictions to guide experiments.  Such a presentation is 
 increasingly valuable as state-of-the-art neutron-scattering instrumentation 
should allow the spin dynamics of real materials to be searched for the discussed unusual features 
with sufficient energy resolution and momentum-space coverage.  

The present analysis does not include the case of frustrated interplanar exchanges for which enhanced quantum fluctuations~\cite{Yildirim1998} should result in decays of comparable magnitude to the purely 2D case. Additionally, ferromagnetic interlayer coupling, while experimentally appealing due to the reduction of the threshold field $H^*$~\cite{mike_unpub}, goes beyond the scope of the present work.  Nonetheless it is expected that the inclusion of any type of increased dimensionality of sufficient strength will act to soften Van Hove singularities without inherently removing decay phenomena.

The landscape of the high-field dynamical structure factor shown in this work is diverse and intriguing.  
Our results, consistent with prior numerical and analytical studies, provide a 
comprehensive illustration of its details. 
Inelastic neutron-scattering measurements on suitable quantum antiferromagnets 
will be important to elucidate the accuracy of the theoretical results presented in this work.  

\begin{acknowledgments}
M.~M. would like to thank N.~B.~Christensen, 
H.~M.~R\o{}nnow, and D.~F.~McMorrow for numerous stimulating discussions.
This work was supported in part by the UROP and SURP programs at UC Irvine (W.~T.~F.), 
graduate fellowship at ILL Grenoble (M.~M.), and the US Department of Energy 
under grants DE-FG02-08ER46544 (M.~M.) 
and DE-FG02-04ER46174 (A.~L.~C.).
\end{acknowledgments}

\appendix

% -------------------------------------------------------------------------
\section{Details of Spin-Wave Theory}
\label{ap1}
% -------------------------------------------------------------------------

The derivation of $1/S$ corrections to spin-wave theory in an applied magnetic field
with non-zero interlayer antiferromagnetic exchange begins with the Heisenberg
Hamiltonian of nearest-neighbor exchange and Zeeman contributions,
        \begin{equation}
                \mathcal{H} = \sum_{\langle ij\rangle}
                        J_{ij} {\bf S}_i\cdot{\bf S}_j  - H \sum_i {S}^{z_0}_i ,
         \end{equation}
Where the factor $g\mu_B$, with $g$ being the gyromagnetic ratio and  $\mu_B$ the
Bohr magneton, 
is included in the applied magnetic field  $H$.  We consider the Heisenberg
antiferromagnet on the
simple tetragonal lattice,
 with nearest-neighbor exchange in-plane coupling $J$ and interlayer coupling $J'$, 
 parametrized by $\alpha = J'/J$, $0\leq\alpha\leq1$. 
 The long-range antiferromagnetic order of spins canted by external field is
described by the 
 same ordering wave-vector $\mathbf{Q} = (\pi,\pi,\pi)$ as in the collinear N\'eel
order at $H = 0$.

Spin components in the laboratory frame $(x_0,y_0,z_0)$ are 
related to that in rotating frame $(x,y,z)$ via the transformation
        %---------------
        \begin{eqnarray}
                S_i^{z_0} & = &   S_i^{z}\sin\theta \nonumber
                                -e^{i\g{Q}\cdot\g{r}_i} S_i^{x}\cos\theta , \\              
                S_i^{x_0} & = &   e^{i\g{Q}\cdot\g{r}_i} S_i^{z}\cos\theta 
                                + S_i^{x}\sin\theta,                \\
                S_i^{y_0} & = &   S_i^{y},\nonumber
\label{frame}        
        \end{eqnarray}
        %---------------
with $\theta$ the canting angle from the spin-flop plane. The
transformed Hamiltonian reads
  %---------------
  \begin{eqnarray}
                \mathcal{H} =  \sum_{\langle ij \rangle}\: J_{ij}\: 
                  \Big(                     S_i^{y}S_j^{y} 
                         - \cos2\theta \big( S_i^{x}S_j^{x} + S_i^{z}S_j^{z} \big) \nonumber \\
                   \hspace{0.5cm}  - e^{i\g{Q}\cdot\g{r}_i}\sin2\theta 
                                       \big( S_i^{x}S_j^{z} - S_i^{z}S_j^{x} \big)    
                  \Big)  \\
                           - H \:\sum_{i}\:                          
                  \big(S_i^{z}
                  \sin\theta  - e^{i\g{Q}\cdot\g{r}_i}S_i^x\cos\theta   
                  \big),\nonumber
        \end{eqnarray}
        %---------------
where $\langle ij \rangle$ designate a single counting of nearest neighbor exchanges.  
Using the usual Holstein-Primakoff transformation\cite{Holstein1940} and keeping up
to quartic terms we arrive at the following terms of order $\mathcal{O}(S^{2-n/2})$:
  %---------------
  \begin{eqnarray}
  % Clasical Energy
    \varepsilon_0 & = & \mathcal{H}_0/N = - JS^2 
                                                                      \left(  (2+\alpha)\cos2\theta
                                 +\sin\theta\frac{H}{JS} \right)   \\[2ex]
%
%
%
    % Linear                                 
    \mathcal{H}_1 & = & \cos\theta\:\sum_i 
                        e^{i\g{Q}\cdot\g{r}_i}\: S_i^x
     \big(H - 4JS\sin\theta\:(2+\alpha) \big)  \\   
%
%
%
   % Quadratic                                 
    \mathcal{H}_2 & = & \sum_i \Big[
                      H\sin\theta\: a_i\da a_i  \nonumber\\
                            && \:+ \: \frac{S}{2} \:
                                                                  \sum_{j(i)} J_{ij}
    \Big(  \cos2\theta  \,\big( a_i\da a_i + a_j\da a_j \big) \\ && 
          +\sin^2\theta  \,\big( a_i\da a_j + a_j\da a_i \big) 
          -\cos^2\theta  \,\big( a_i\da a_j\da + a_j a_i \big) \nonumber
    \Big)  \Big]  \nonumber\\
%
%
% 
   % Cubic
    \mathcal{H}_3 & = & \sqrt{\frac{S}{2}} \sum_i
                                                                                     e^{i\g{Q}\cdot\g{r}_i} \Bigg[
             \left(\frac{H\cos\theta}{4S}-\frac{J(2+\alpha)\sin2\theta}{2}\right)\\
           &&  \times \big(a_i\da a_i\da a_i +  a_i\da a_i a_i\big)        
+ \sin2\theta \sum_{j(i)} 
             J_{ij} \big(a_i\da + a_i\big) a_j\da a_j                                          \Bigg] \nonumber\\                                
  % Quartic
    \mathcal{H}_4 & = & \sum_{i,\:j(i)} J_{ij}\Bigg[ 
      \frac{1}{4}\cos^2\theta \,\big( (n_i+n_j)\, a_ia_j + \mbox{h.c.}\big)\\
     && -\frac{1}{4}\sin^2\theta\, \big( a_i\da(n_i+n_j)\, a_j + \mbox{h.c.}  \big)
     -\cos2\theta \, n_i n_j \Bigg],\nonumber
        \end{eqnarray} 
        %---------------     
where $j(i)$ refers to a site $j$ that is nearest neighbor of $i$.

\subsection{Harmonic approximation}
First, we minimize $\varepsilon_0$ to obtain the classical
value of the canting angle
        %---------------
        \begin{eqnarray}
         \sin\theta = \frac{H}{4JS(2+\alpha)} \equiv \frac{H}{H_s}
  \end{eqnarray} 
        %---------------   
where $H_s\!=\!4JS(2+\alpha)$ is the saturation field where the spin structure
reaches fully polarized state. 
Then, the Fourier transformation of $\mathcal{H}_2$ leads to the usual
        %---------------
        \begin{eqnarray}
         \mathcal{H}_2 = \sum_{\g{k}}\bigg(
                                         A_{\g{k}} a\ka\da a\ka - \frac{1}{2} B_{\g{k}} 
                                         \left(a\ka\da a_{-\g{k}}\da + a\ka a_{-\g{k}}\right)\bigg)\, , \ \ \ \ \ \ 
  \end{eqnarray} 
        %---------------     
where the coefficients $A\ka$ and $B\ka$ read
        %---------------
        \begin{align}  &A\ka =  2JS(2+\alpha) \left( 1 
                                            + \sin^2\theta\:\bar{\gamma}{\ka}\right) {\;\;\;}   \\
        & B\ka  =  2JS(2+\alpha) \cos^2\theta \:\bar{\gamma}{\ka}\nonumber
  \end{align} 
        %---------------     
and where the lattice harmonics are defined as $\bar{\gamma}\ka = (\gamma_{xy} +
\alpha \gamma_{z})/(2+\alpha)$        with $\gamma_{xy} = (\cos k_x+\cos k_y)$ and
$\gamma_z = \cos k_z$. 
Upon the Bogolioubov transformation and substitution of the classical canting angle, 
we obtain the dispersion relation,
        %---------------
        \begin{align}
                &\varepsilon\ka \equiv 2JS(2+\alpha) \omega\ka,   \\
                &\omega\ka = 
                \sqrt{(1+\bar{\gamma}{\ka})(1-\cos2\theta\:\bar{\gamma}\ka)} \nonumber  \:
\label{LSWT}
  \end{align} 
        %---------------  
and the parameters of the transformation
        %---------------
        \begin{eqnarray}
          u^2\ka,v^2\ka &=& \frac{A\ka \pm \varepsilon\ka}{2\varepsilon\ka},
          \hspace{1cm} u\ka  v\ka  = \frac{B\ka}{2\varepsilon\ka}.
  \end{eqnarray} 
        %---------------  

\subsection{Mean-Field decoupling}

The quartic Hamiltonian $\mathcal{H}_4$ contains four-boson terms that can be treated 
using mean-field Hartree-Fock decoupling. To do so, we introduce the following
averages,
        %---------------
        \begin{align}
                &n = \langle a_i\da a_i \rangle = \sumk v\ka^2 \:\:,\nonumber \\
                &\delta   =  \langle a_i    a_i \rangle = \sumk u\ka v\ka  \nonumber\:\:,  \\
                &m_{xy} = \langle a_i\da a_j \rangle_{xy} 
                             = \sumk \frac{\gamma_{xy}}{2} v\ka^2 \:\:,  \\
                &m_{z}   = \langle a_i\da a_j \rangle_{z}  
                             = \sumk \gamma_{z} v\ka^2 \nonumber\:\:,  \\
                &\Delta_{xy} = \langle a_i a_j \rangle_{xy} 
                             = \sumk \frac{\gamma_{xy}}{2} u\ka v\ka \:\:,  & \nonumber\\
                &\Delta_{z}   = \langle a_i a_j \rangle_{z}  
                             = \sumk \gamma_{z} u\ka v\ka \nonumber\:\:. 
        \end{align} 
        %---------------
In zero magnetic field, anomalous averages vanish so that $\delta = m_{xy} = m_z =
0$ while the average $n$ determines the sublattice magnetization associated with the
linear spin-wave theory 
$\langle S^z \rangle = S - n$. Evaluation of the Hartree-Fock averages in finite
magnetic field is 
performed using default global adaptive integration methods via Mathematica with an
accuracy 
goal of greater than 5 digits.

 Using the method of Oguchi\cite{Oguchi1960} to treat the quartic term, we obtain
the quadratic Hamiltonian,
        %---------------
        \begin{eqnarray}
        \langle \mathcal{H}_4 \rangle_1 & = &
          \sumk \delta A^{(4)}\ka a\ka\da a\ka-\frac{\delta B^{(4)}\ka}{2}\left(
                 a\ka\da a_{\g{-k}}\da + a\ka a_{\g{-k}}\right),  \hspace{1cm}\\
                 \mbox{ where} \nonumber\\
\label{dimensionless}
         %----         
          \delta A^{(4)}\ka & = & 2J (2+\alpha) \Big[
           \Delta\cos^2\theta - n\cos2\theta - m \sin^2\theta \nonumber\\
           &&\hspace{.4cm} +\frac{\bar{\gamma} \ka}{2} \left(\delta \cos^2\theta -
2n\sin^2\theta \right)
           -\bar{\gamma}_m \cos2\theta 
          \Big]         \nonumber \\
         %----         
          \delta B^{(4)}\ka & = & 2J(2+\alpha) \Bigg[
           \frac{1}{2}\left(\Delta\sin^2\theta - m \cos^2\theta\right) \\
           &&        \hspace{.4cm}   +\frac{\bar{\gamma}\ka}{2} ( \delta \sin^2\theta - 2
n\cos^2\theta )
           +\bar{\gamma}_{\Delta} \cos2\theta 
          \Bigg]          \nonumber
        \end{eqnarray} 
        %---------------
using the definitions
        %---------------
        \begin{alignat}{2}
         &m  =  \frac{2m_{xy}+\alpha m_{z}}{2+\alpha},
         &&\hspace{.3cm}\Delta = \frac{2\Delta_{xy}+\alpha\Delta_{z}}{2+\alpha},  \\
         &\bar{\gamma}_m  =  \frac{m_{xy}\gamma_{xy}
                                          + \alpha m_{z}\gamma_z}{2+\alpha}, 
         &&\hspace{.3cm}\bar{\gamma}_\Delta  =  \frac{\Delta_{xy}\gamma_{xy}
                                          + \alpha\Delta_{z}\gamma_z}{2+\alpha}.\nonumber
        \end{alignat}
        %---------------        
The corresponding correction to the dispersion reads
        %---------------
        \begin{eqnarray}
        \delta\varepsilon\ka^{(1)} & = & 2J (2+\alpha) \\
        &&        \times \frac{1}{\omega\ka}\Big[\big(1+\bar{\gamma}\ka\sin^2\theta\big)
                \delta \tilde{A}\ka^{(4)}
                - {\gamma}\ka\cos^2\theta \delta \tilde{B}\ka^{(4)} \Big] \nonumber
\label{cor1}
  \end{eqnarray}
  %---------------        
where $\tilde{A}\ka$  and $\tilde{B}\ka$ are dimensionless expressions in the brackets of (\ref{dimensionless}).  

%---------------------------------------------------------------
\subsection{Angle Renormalization}
        
In contrast to the zero-magnetic field case, the $1/S$ expansion of the present
Hamiltonian 
contains cubic terms. The Hartree-Fock decoupling of the cubic term $\mathcal{H}_3$ 
is responsible for a renormalization of the canting angle and yields
        %---------------
        \begin{eqnarray}
          \langle \mathcal{H}_3 \rangle & = & \sqrt{2S}\, J(2+\alpha)\\
&&          \times\sin2\theta        (n-\Delta-m)\left(a_{\g{Q}}\da+a_{\g{Q}} 
\right).\nonumber
  \end{eqnarray}
  %---------------        
The value of the renormalized angle is obtained from the cancellation of the linear
term  
$\mathcal{H}_1 
+ \langle\mathcal{H}_3\rangle$ so that
        %---------------
        \begin{eqnarray}
    \sin\tilde{\theta} & = & \sin\theta \left[ 1 + \frac{n-m-\Delta}{S} \right],
  \end{eqnarray}
  %---------------        
and the corresponding correction to the dispersion is therefore
        %---------------
        \begin{eqnarray}
        \delta\varepsilon\ka^{(2)} & = & 4J (2+\alpha)\sin^2\theta \\ 
&&                \times \frac{(\Delta + m - n)}{\omega\ka}
                \Big[ 1 - \bar{\gamma}\ka^2-\bar{\gamma}\ka\cos^2\theta \Big].\nonumber
\label{cor2}
  \end{eqnarray}
  %---------------        
  
%---------------------------------------------------------------
\subsection{Cubic Vertices}

Interaction vertices are obtained upon Fourier transformation of the fluctuating
part of $\mathcal{H}_3$, 
%
%        %---------------
\begin{eqnarray}
\mathcal{H}_3 & = & \sqrt{2S}\, J(2+\alpha) \\
&&                        \times \sin 2\theta
                \sum_{\g{k},\g{q}} \bar{\gamma}\ka 
                \left( a\ka\da a_{\g{q}}\da a_{\g{k-q+Q}} 
                + \mbox{\small h.c} \right)\nonumber
        \end{eqnarray}                        
%  %---------------        
leading to the following cubic decay 
and source vertices, $\Phi_{31}(\g{k},\g{q})$ and $\Phi_{32}(\g{k},\g{q})$,  of the
form
$\Phi_{31,32}(\g{k},\g{q}) =  -\sqrt{2S}\, J (2+\alpha)
\sin2\theta \:\tilde{\Phi}_{31,32}(\g{k},\g{q})$, where
        %---------------
        \begin{eqnarray}
\widetilde{\Phi}_{31}  (\mathbf{k},\mathbf{q}) & = &
\bar{\gamma}_{\mathbf{k}} ( u_{\mathbf{k}}+v_{\mathbf{k}}) 
(u_{\mathbf{q}} v_{\mathbf{k-q+Q}} + v_{\mathbf{q}} u_{\mathbf{k-q+Q}}  )
\label{eq:gam1}    \nonumber      \\ 
& + & \bar{\gamma}_{\mathbf{q}} (u_{\mathbf{q}}+v_{\mathbf{q}}) 
(u_{\mathbf{k}}u_{\mathbf{k-q+Q}} + v_{\mathbf{k}}v_{\mathbf{k-q+Q}}  ) 
\nonumber \\
& + & \bar{\gamma}_{\mathbf{k-q+Q}}(u_{\mathbf{k-q+Q}}+v_{\mathbf{k-q+Q}}) 
(u_{\mathbf{k}}u_{\mathbf{q}} + v_{\mathbf{k}}v_{\mathbf{q}}  ) \ ,
\nonumber \\
\widetilde{\Phi}_{32}  (\mathbf{k},\mathbf{q}) & = &
\bar{\gamma}_{\mathbf{k}} ( u_{\mathbf{k}}+v_{\mathbf{k}}) 
(u_{\mathbf{q}} v_{\mathbf{k-q+Q}} + v_{\mathbf{q}} u_{\mathbf{k-q+Q}}  )
 \\ 
& + & \bar{\gamma}_{\mathbf{q}} (u_{\mathbf{q}}+v_{\mathbf{q}}) 
(u_{\mathbf{k}}v_{\mathbf{k-q+Q}} + v_{\mathbf{k}}u_{\mathbf{k-q+Q}}  ) 
\nonumber \\
& + & \bar{\gamma}_{\mathbf{k-q+Q}}(u_{\mathbf{k-q+Q}}+v_{\mathbf{k-q+Q}}) 
(u_{\mathbf{k}}v_{\mathbf{q}} + v_{\mathbf{k}}u_{\mathbf{q}}  ) \ .
\nonumber 
        \end{eqnarray}                        
  %---------------        
The corresponding self-energy corrections therefore read
        %---------------
        \begin{eqnarray}
        \Sigma_{31}(\mathbf{k},\omega) & = & 2J(2+\alpha)\cos^2\theta\sin^2\theta \\
&&        \times  \nonumber  \sum_{\g{q}}
            \frac{|\tilde{\Phi}_{31}(\g{k},\g{q})|^2}{\omega - \omega_{\g{q}} 
           - \omega_{\g{k-q+Q}} + i 0^+} \\ 
  \Sigma_{32}(\mathbf{k},\omega) & = & -2J(2+\alpha) \cos^2\theta\sin^2\theta\\
&&        \times 
        \sum_{\g{q}}
        \frac{|\tilde{\Phi}_{32}(\g{k},\g{q})|^2}{\omega + \omega_{\g{q}} 
           + \omega_{\g{Q-k-q}}  - i 0^+}           \nonumber   
        \end{eqnarray}                        
  %---------------  
where $0^+$ is a small positive number.  In taking these integrals, $0^+$ was taken
as $10^{-2}$.   

 Then the  \emph{on-shell} correction to the spectrum is, 
        %---------------
        \begin{eqnarray}
 \delta\varepsilon^{(3)}\ka  &=&  J\left(1+\frac{\alpha}{2}\right)
        \sin^22\theta
          \sum_{\g{q}} \Bigg(
            \frac{|\tilde{\Phi}_{31}(\g{k},\g{q})|^2}{\omega\ka - \omega_{\g{q}} 
           - \omega_{\g{k-q+Q}} + i0^+}\nonumber \\
&&        \phantom{ J\left(1+\frac{\alpha}{2}\right)} 
                -\frac{|\tilde{\Phi}_{32}(\g{k},\g{q})|^2}{\omega\ka + \omega_{\g{q}} 
           + \omega_{\g{Q-k-q}} - i0^+} \Bigg)
\label{cor3}
        \end{eqnarray}
  %---------------  
so that combining all the contributions together we obtain the renormalized
spin-wave dispersion
        %---------------
        \begin{eqnarray}
          \bar{\varepsilon}\ka & = & \varepsilon\ka + \delta\varepsilon^{(1)}\ka
          + \delta\varepsilon^{(2)}\ka + \delta\varepsilon^{(3)}\ka,
  \end{eqnarray}                        
  %---------------  
where $\varepsilon\ka$ is the linear spin-wave theory energy, given in Eq.~(\ref{LSWT}),
 $\delta\varepsilon^{(1)}\ka$ is the correction due to Hartree-Fock decoupling of the
 quartic terms from Eq.~(\ref{cor1}), 
 $\delta\varepsilon^{(2)}\ka$  is the correction due to angle renormalization in Eq.~(\ref{cor2}), 
 and $\delta\varepsilon^{(3)}\ka$  is the correction due to cubic vertices from Eq.~(\ref{cor3}).  
  
%---------------------------------------------------------------

%---------------------------------------------------------------

\begin{thebibliography}{99}
%---------------------------------------------------------------

\bibitem{Manousakis}
E. Manousakis, Rev. Mod. Phys. {\bf 63}, 1 (1991).

\bibitem{Mattis} D. C. Mattis, {\it The Theory of magnetism made simple: 
An Introduction to Physical Concepts and to Some Useful Mathematical Methods}, 
World Scientific (2006).

\bibitem{Diep}  {\it Frustrated spin systems},
edited by H. T. Diep, 
World-Scientific, Singapore (2003).

\bibitem{Johnston} D. C. Johnston, in {\it Handbook of Magnetic Materials},
  vol. 10, (ed. by K. H. J. Buschow, Elsevier Science, North Holland, 1997).

\bibitem{Balents} L. Balents, Nature {\bf 464}, 199 (2010).

\bibitem{Giamarchi08} T. Giamarchi, C. R\"{u}egg, and O. Tchernyshyov,
Nature Physics, {\bf 4}, 198 (2008).

\bibitem{Stone06} M. B. Stone, I. A. Zaliznyak, T. Hong, D. H. Reich
  and C. L. Broholm,  Nature {\bf 440}, 187 (2006). 

\bibitem{Woodward02}
F. M. Woodward, A. S. Albrecht, C. M. Wynn, C. P. Landee, and M. M. Turnbull,
Phys. Rev. B {\bf 65}, 144412 (2002).

\bibitem{Lancaster07}
T. Lancaster, S. J. Blundell, M. L. Brooks, P. J. Baker, F. L. Pratt,
J. L. Manson, M. M. Conner, F. Xiao, C. P. Landee, F. A. Chaves, S. Soriano,
M. A. Novak, T. Papageorgiou, A. Bianchi, T. Herrmannsdorfer, J. Wosnitza,
and J. A. Schlueter,
Phys. Rev. B {\bf 75}, 094421 (2007).

\bibitem{Coomer07}
F. C. Coomer, V. Bondah-Jagalu, K. J. Grant,  A. Harrison,
G. J. McIntyre, H. M. R\o{}nnow,
R. Feyerherm, T. Wand, and M. Meissner, D. Visser, and D. F. McMorrow,
Phys. Rev. B \textbf{75}, 094424 (2007).

\bibitem{Xiao09} F. Xiao, F. M. Woodward, C. P. Landee, M. M. Turnbull, 
C. Mielke, N. Harrison, T. Lancaster, S. J. Blundell, P. J. Baker, P. Babkevich, and F. L. Pratt
Phys. Rev. B {\bf 79}, 134412 (2009).

\bibitem{Tsyrulin09}
N. Tsyrulin, T. Pardini, R. R. P. Singh, F. Xiao, P. Link, A. Schneidewind, A. Hiess, C. P. Landee,
M. M. Turnbull, and M. Kenzelmann, Phys. Rev. Lett. {\bf 102}, 197201 (2009).

\bibitem{Tsyrulin10} N. Tsyrulin, F. Xiao, A. Schneidewind, P. Link, H. M. R\o{}nnow,
 J. Gavilano, C. P. Landee, M. M. Turnbull, and M. Kenzelmann, Phys. Rev. B {\bf 81}, 134409 (2010).

\bibitem{Ollivier2010} J. Ollivier, H. Mutka, and L. Didier, Neutron News {\bf 21}, 22 (2010), 
(www.tandfonline.com/doi/pdf/10.1080 /10448631003757573).

\bibitem{Bewley2011} R. Bewley, J. Taylor, and S. Bennington, 
Nucl. Instr. and Meth. A {\bf 637}, 128 (2011).

\bibitem{Ehlers2011} G. Ehlers, A. A. Podlesnyak, J. L. Niedziela, E. B. Iverson, and 
P. E. Sokol, Rev. Sci. Instrum. {\bf 82}, 085108 (2011).

\bibitem{J1J2J3} P. Chandra, P. Coleman and A. I. Larkin, J. Phys. Condens. Matter {\bf 2}, 7933 (1990)

\bibitem{Henley} C. L. Henley, Phys. Rev. Lett. {\bf 62}, 2056 (1989).

\bibitem{Chernyshev06}
A. L. Chernyshev and M. E. Zhitomirsky, 
Phys. Rev. Lett. {\bf 97}, 207202 (2006). 

\bibitem{triangle} A. L. Chernyshev and M. E. Zhitomirsky, 
Phys. Rev. B {\bf 79}, 144416 (2009). 

\bibitem{field}
M. E. Zhitomirsky and A. L. Chernyshev,
Phys. Rev. Lett. {\bf 82}, 4536 (1999).

\bibitem{Mourigal2010} M. Mourigal, M.E. Zhitomirsky, A. L. Chernyshev,
Phys. Rev. B {\bf 82}, 144402 (2010). 

\bibitem{Kreisel08}
A. Kreisel, F. Sauli, N. Hasselmann, and P. Kopietz,
Phys. Rev. B {\bf 78}, 035127 (2008).

\bibitem{Syromyat09}
A. V. Syromyatnikov,
Phys. Rev. B {\bf 79}, 054413 (2009).

\bibitem{Olav08} 
O. F. Sylju\aa sen, Phys. Rev. B {\bf 78}, 180413 (2008). 

\bibitem{Lauchli09} A. L\"{u}scher and A. L\"{a}uchli, Phys. Rev. B {\bf 79}, 195102 (2009). 

\bibitem{Masuda10}
T. Masuda, S. Kitaoka, S. Takamizawa, N. Metoki, K. Kaneko,
K. C. Rule, K. Kiefer, H. Manaka, and H. Nojiri, Phys. Rev. B \textbf{81}, 100402(R) (2010).

\bibitem{Kohama2011} Y. Kohama, A. Sologubenko, N. R. Dilley, V. S. Zapf, M. Jaime, 
J. Mydosh, A. Paduan-Filho, K. Al-Hassanieh, P. Sengupta, S. Gangadharaiah, A. L. Chernyshev, and C. D. Batista, 
Phys. Rev. Lett. {\bf 106}, 037203 (2011).

\bibitem{sasha_unpub} A. L. Chernyshev, unpublished.

\bibitem{Nikuni98}
M. E. Zhitomirksy and T. Nikuni,
Phys. Rev. B \textbf{57}, 5013 (1998).

\bibitem{footnote} 
We note that Fig.~\ref{regions} shows only the boundaries given by the decay threshold 
determined by the decay into a pair of magnons with the same momenta. Generally, such a boundary is
the leading one, with some further extensions due to other processes that may occur at higher fields.
For details, see Ref.~\onlinecite{Mourigal2010}.

\bibitem{footnote1}  Some singularities, even after being regularized, may remain parametrically enhanced 
 relative to the background 
 and correspond to selected spots in the momentum space where broadening is large.\cite{Mourigal2010}

\bibitem{Squires} 
G. L. Squires, {\it Introduction to the Theory of Thermal Neutron Scattering}, (Dover, New York, 1996).

\bibitem{Birgeneau}
I. U. Heilmann, J. K. Kjems, Y. Endoh, G. F. Reiter,
G. Shirane, and R. J. Birgeneau,
Phys. Rev. B \textbf{24}, 3939 (1981).

\bibitem{Holstein1940}
T. Holstein and H. Primakoff,
Phys. Rev. \textbf{58}, 1098 (1940).

\bibitem{Oguchi1960}
T. Oguchi, Phys. Rev. \textbf{117}, 117 (1960).

\bibitem{Yildirim1998}
T. Yildirim, A. B. Harris, and E. F. Shender, Phys. Rev. B \textbf{58}, 3144 (1998).

\bibitem{mike_unpub} M. E. Zhitomirsky, M. Mourigal, and A. L. Chernyshev, unpublished.


\end{thebibliography}
\end{document}